# A Manifesto for Modern Fog and Edge Computing: Vision, New Paradigms, Opportunities, and Future Directions

Sukhpal Singh Gill

School of Electronic Engineering and Computer Science, Queen Mary University of London, UK

**Correspondence:** Sukhpal Singh Gill, School of Electronic Engineering and Computer Science, Queen Mary University of London, Mile End Rd, Bethnal Green, London E1 4NS, UK.

Email: s.s.gill@qmul.ac.uk, https://orcid.org/0000-0002-3913-0369

## Abstract

The advancements in the use of Internet of Things (IoT) devices is increasing continuously and generating huge amounts of data in a fast manner. Cloud computing is an important paradigm which processes and manages user data effectively. Further, fog and edge computing paradigms are introduced to improve user service by reducing latency and response time. This chapter presents a manifesto for modern fog and edge computing systems based on the current research trends. Further, architectures and applications of fog and edge computing are explained. Moreover, research opportunities and promising future directions are presented with respect to the new paradigms, which will be helpful for practitioners, researchers and academicians to continue their research.

**Keywords:** IoT, Edge Computing, Cloud Computing, Manifesto, Architecture, Fog Computing

## 1. Introduction

In modern era, the utilization of Internet of Things (IoT) applications is growing exponentially and generating a large amount of data [1-2]. Prominent Tech giants (Facebook, Microsoft, Google, Apple and Amazon) are using massive Cloud Datacenters (CDCs) to provide the effective user services to process this data in a proficient and trustworthy manner [3-4]. Some critical and deadline oriented IoT applications such as traffic management, military or healthcare, needs to process their requests with minimum latency and response time [5-6]. To solve this problem, the two advanced paradigms, fog and edge computing are introduced to process and manage data quickly [7-8]. Fog devices aid cloud to process the small user requests without sending it to cloud while edge devices provide an innovative service to execute user request at edge device [9]. The research in both fog and edge computing is growing day by day since they emerged as groundbreaking computing paradigms [10]. Therefore, there is a need to assess the existing research related to emerging paradigms to find out the possible future directions and research opportunities existing in this field.

## 1.1 Motivation

Research in the area of fog and edge computing is growing day by day. There are a few review and survey articles in the literature [1, 2, 3, 4, 5, 6, 7] but the research has purposefully grown in modern fog and edge computing. It is important to evaluate, advance and upgrade the existing research of modern fog and edge computing. This work expands the existing research and presents a fresh manifesto on modern edge and fog computing to identify the further opportunities and future directions for emerging paradigms.

## 1.2 Contributions

In this work, we have summarized the existing research in terms of edge and fog computing and presented the relationship of fog and edge computing with new paradigms. A thorough investigation has been carried out to uncover the existing challenges and research questions. Research opportunities in the area of edge and fog computing are presented.

**1.3 Chapter Structure**

The reminder of the chapter is organized as follows: The background of edge and fog computing is presented in Section 2. Section 3 describes the applications of edge and fog computing. Architecture types of edge and fog computing are discussed in Section 4. Section 5 discusses new paradigms and Section 6 presents the research opportunities and future directions. Lastly, Section 7 summarizes the chapter.

**2. Background: Fog and Edge Computing**

The concept of fog computing was emerged in the year 2009 and was adopted in the year 2011 as an assistance to cloud computing [11] and manages the extensive data incoming from IoT devices continuously. Fog computing servers are deployed among edge devices and CDCs to process the small jobs quickly instead of sending to the CDC [12]. Further, edge computing is emerged in the 2010 and adopted in the year 2012, which helps users to process their requests at their edge devices instead of sending to fog or cloud server [13]. Both fog and edge computing optimizes energy, execution cost and time while maintaining the quality of service at required level. Figure 1 shows the basic model of fog and edge computing, which contains four main layers: cloud, fog, edge and IoT. Cloud layer consist of data centers and process the user requests which needs large processing capability to manage big chunk of data. Fog layer contains the fog devices (servers or nodes) to process the small user requests to reduce latency and response time and it needs less processing capability as compared to cloud computing. Edge layer contains edge devices to process user request at edge device despite of sending it to fog or cloud server. IoT layer comprise of IoT devices related to the various applications such as healthcare, agriculture, traffic monitoring, weather forecasting or other sensors related to particular application.

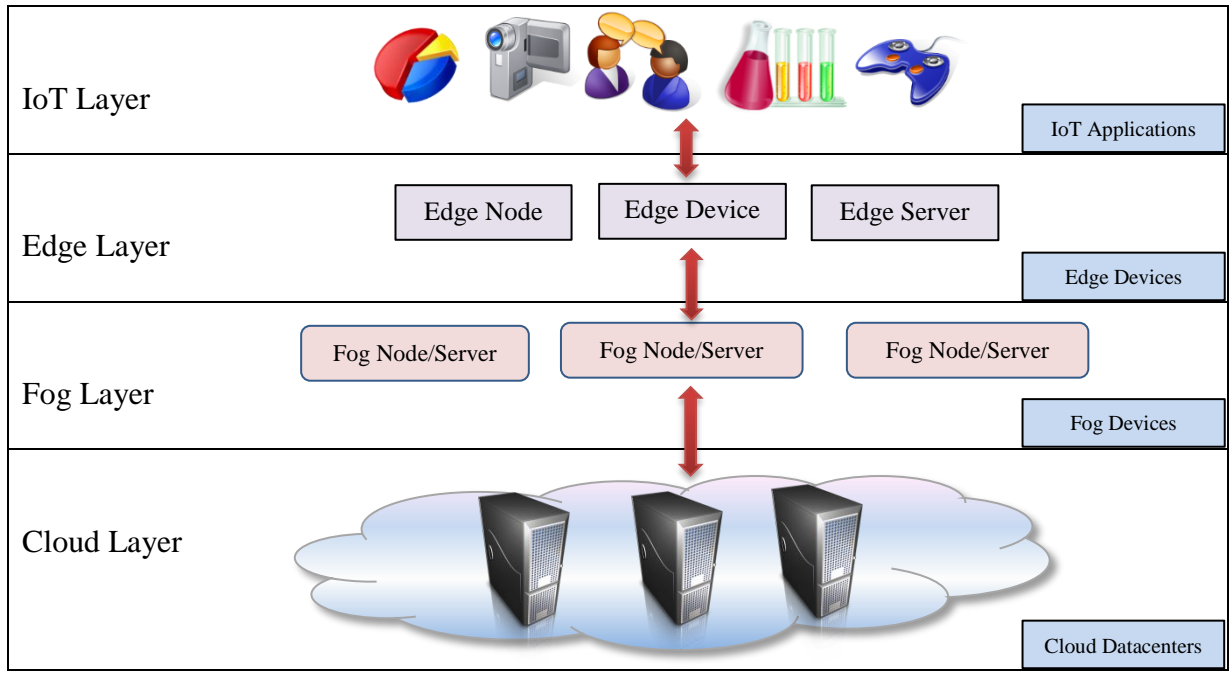

Fig. 1. Fog and Edge Computing Model

## 3. IoT Applications

IoT applications such as weather forecasting, water supply control, traffic monitoring, smart home & city, agriculture and healthcare are utilizing fog and edge computing paradigms to process user data to reduce latency and response time. Figure 2 shows key IoT applications.

### 3.1 Healthcare

Healthcare is an important IoT application which is developed for different diseases such as cardiac arrest, diabetic, cancer, COVID-19 or pneumonia [14]. For example, an integrated fog and edge computing environment is used to diagnose the status of heart patient automatically using various medical sensors [15]. Further, the current healthcare service can be improved by using next generation technologies such as artificial intelligence or virtual reality to combat future pandemics [16].

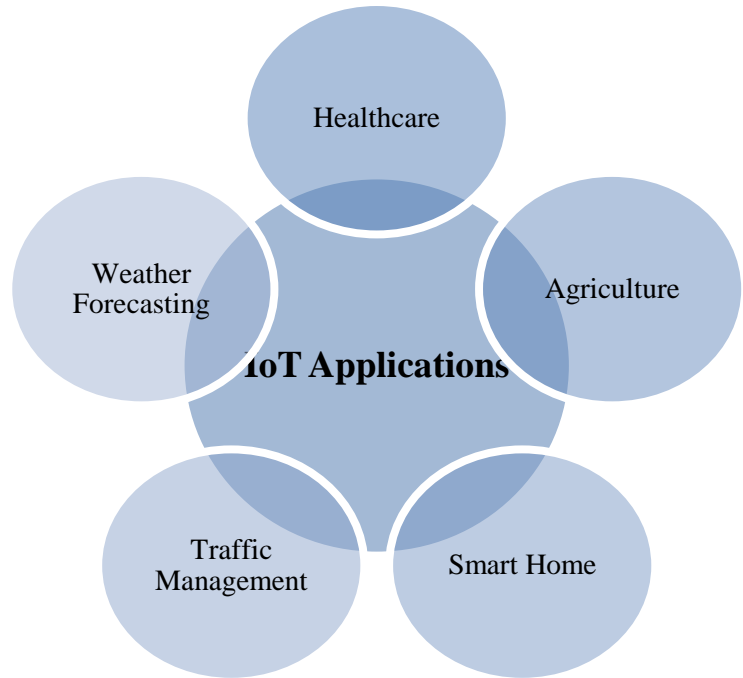

Fig. 2. IoT Applications

## 3.2 Agriculture

Agriculture sector is using advanced technology to process various types of agriculture-related data for the prediction of different parameters such as production rate, water level or crop quality [17]. For example, cloud based agriculture system is developed to predict the agriculture status automatically based on the information provided by various IoT or edge devices [18]. Further, a mobile application is developed to manage the enormous volume of data and provide the required information to the farmers at their edge device to enable smart farming.

## 3.3 Smart Home

Electricity is an important source of power, which must be used wisely [19]. With the development smart home, owner can control their home appliances from mobile to optimize the use of energy and provide the required security by deploying cameras. For example, cloud and fog computing based resource management technique is developed to control the edge devices using mobile application and controls fans, lights, cameras, room temperature and voltage using sensors related to various home appliances [19].

**3.4 Traffic Management**

IoT is playing an important role for managing traffic effectively using various sensors and actuators. For example, a smart transportation system is developed for the detection of potholes using IoT devices [20]. Further, various machine learning techniques were used to evaluate its performance in terms of performance parameters. Furthermore, fog and edge computing paradigms can be used to process data quickly for the earlier notification of potholes to avoid accidents.

**3.5 Weather Forecasting**

IoT and cloud computing can contribute towards weather and climate forecasting and science by facilitating observations empowered by IoT devices [21]. Traditionally measurements of weather parameters such as air pollution, humidity and others are observed, saved and displayed to the public and are also used by the researchers to improve the science behind the natural phenomena [22]. A sensor-based IoT system can be used for these observations which can send the data to the cloud.

**4. Type of Architectures**

Basically, there are three different types of architectures for fog and edge computing [23-25]: centralized, decentralized and hierarchical as shown in Figure 3. There is one chief controller in the *centralized* architecture, which controls all the internal activities such as task scheduling, provisioning of resources, scheduling of resources and its execution. There are two different levels of schedulers in *hierarchical* architecture: upper and lower level. Upper level is a chief broker which manages all the lower-level schedulers while lower level schedulers perform various internal activities independently. *Decentralized* architecture allocates the resources to various user requests based on their specific necessities to complete their task, which also provides a high level of scalability.

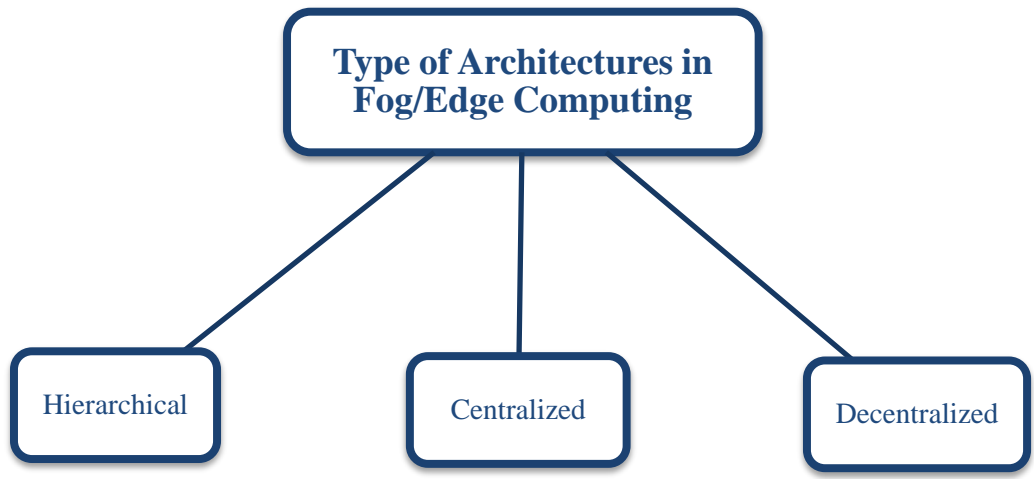

Fig. 3. Type of Architectures in Fog and Edge Computing

## 5. Relevant Issues and Paradigms: A Vision

In this section, a future vision is presented while relevant issues and paradigms are discussed based on the existing research of fog and edge computing. Figure 4 shows the relevant issues and paradigms for fog and edge computing.

### 5.1 Service Level Agreement (SLA) and Quality of Service (QoS)

QoS is a significant measurement parameter, which is used to measure the level of cloud, fog and edge service [1]. QoS parameter can be reliability, security, availability, latency, elasticity, scalability, cost and execution or response time [2]. SLA is an authorized contract, which is assured between cloud provider and end user based on different quality of service requirements and it also contains some SLA deviation based on their negotiation [26-27]. In case of SLA violation, a penalty can be issued to cloud provider or compensation can be given to end user [28-29]. Further, fulfilment of SLA and attainment of required QoS can improve the fog and edge service and increase the system performance.

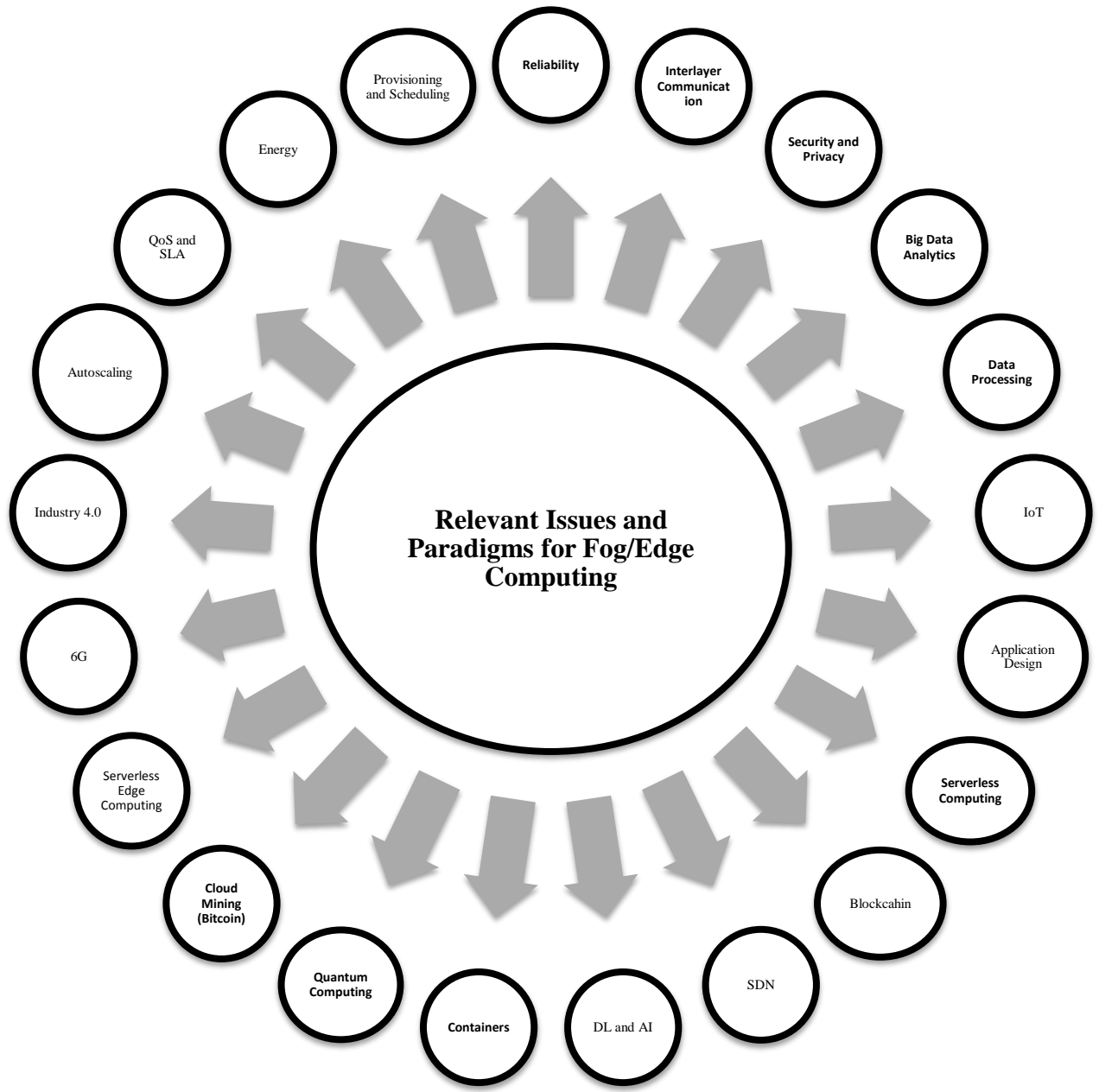

Fig. 4. Relevant Issues and Paradigms for Fog and Edge Computing

## 5.2 Energy Efficiency and Sustainability

There is a need to run the cloud data centers continually to deliver the cloud service in a proficient way, which uses enormous volume of energy for computation and cooling [30-31]. Further, the utilization of large amount of CDC and utilization of massive chunks of energy/power is making an impact on the environment, which leads to climate change by producing carbon footprints [32-33]. To provide the sustainable and energy-efficient fog and edge service, there is a requirement to develop effective energy-efficient techniques for the management of fog/edge resources using machine learning, deep learning or Artificial Intelligence (AI) [34].

## 5.3 Provisioning and Scheduling of Resources

Resource management for fog and edge computing comprises of three main tasks such as provisioning, scheduling and execution of fog/edge resources [35-36]. Resource provisioning is defined as the filtering of best matched resources based on the QoS requirements of the requests [37]. Further, resource scheduler manages the provisioned resources using different scheduling policies based on optimization algorithms, machine learning or AI techniques to optimize the system performance [38]. Finally, resource manager executes the user workloads on scheduled resources.

## 5.4 Fault Tolerance (Reliability)

Fault tolerance is very important concept to maintain the reliability and performance of the system [39]. Fog and edge computing paradigms provide an effective fault tolerant service to run the system continually without any interruption by fixing the faults dynamically [40]. Further, there is a need to find out the software, network or hardware failure and their reasons and fixes to improve the reliability of service [41]. Furthermore, effective fault tolerant techniques can help system to achieve required objective under predefined conditions in a particular time period.

## 5.5 Interlayer Communication

Interlayer communication plays an important role for effective resource management to enable coordination amongst various tasks and subtasks going inside resource manager to perform resource provisioning and scheduling [42]. In fog-edge computing model, there is a requirement of an effective communication among various layers such as cloud, fog, edge and IoT as shown in Figure 1, which can improve the quality of service in terms response time and latency [43].

## 5.6 Security and Privacy

Security and privacy are two important challenges in computing system while delivering services to end users [44]. Privacy-aware fog and edge computing system allow authentic users to access their data using various IoT applications [45]. Further, security plays an important role to provide the safe and secure service to user without any interruption [46]. There is a need of proactive privacy and security-aware systems to protect user data and handles security attacks such as Denial of Service (DoS), Distributed-DoS(DDoS), Remote to Local (R2L), User to Root (U2R) and probing [47] [48].

## 5.7 Big Data Analytics

The growth in connected IoT devices is increasing continuously and generating huge amounts of data in a fast manner, which is called big data [49]. So, big data analytics is an effective way of processing and managing of large amount of data to identify the retrieve information from knowledgebase using innovative data management mechanisms [50]. Fog and edge computing paradigms can be used for proficient big data analytics to improve data processing [51].

## 5.8 Internet of Things (IoT)

IoT applications such as weather forecasting, water supply control, traffic monitoring, smart home & city, agriculture and healthcare are utilizing fog and edge computing paradigms to process user data to reduce latency and response time [52-53]. Fog and edge computing help critical IoT applications to maintain the SLA during the processing of user requests in an efficient manner before their deadline [54-55].

## 5.9 Data Processing

Fog and edge computing helps to gather, process and handle the user data in near proximity of edge devices [56]. Due to the advancement in IoT devices, the generation is increasing quickly, which needs to manage in a better way for effective utilization of storage. With the assistance of effective data processing mechanisms, the user data can be manages efficiently and retrieves the requisite information from the enormous volume of data using fog and edge computing [57].

## 5.10 Application Design

The design of an IoT application is very important for providing effective services using fog and edge computing [58]. Application can be designed using two different ways: monolithic or microservices [59]. Monolithic application is developed as a distinct element, while microservices are presented officially with business-oriented APIs and are very effective in offering scalability due to less coupling.

## 5.11 Serverless Computing

This concept enables the easy and faster development of IoT applications by eradicating the necessity to maintain real infrastructure [60]. Further, Serverless computing or Function as a Service (FaaS) runs the code as independent functions by automatic provisioning of resources and helps to manage the required resources dynamically [61]. Furthermore, execution of independent functions improves the scalability of infrastructure at runtime [62].

## 5.12 Blockchain

This is a new concept of storing data in the form of chain of blocks to increase the security of data [63]. Mostly, ledger are using for transactions to secure data in Blockchain [64]. Further, Blockchain technology can be used to protect the data coming from IoT applications [71].

## 5.13 Software-Defined Network (SDN)

The concept of virtualization can be used for fog and edge computing to provide security and offers virtualization in an efficient manner [65]. Software-Defined Network (SDN) based fog and edge computing can decline energy consumption and increases the utilization of network while processing the data coming from IoT applications [66].

## 5.14 Deep Learning (DL) and Artificial Intelligence (AI)

Artificial Intelligence (AI) is an important branch of computer science, which is working like humans and helps to run various systems automatically [1]. Latest AI or deep learning techniques are effective in managing the resources for fog and edge computing while maintaining the SLA during execution [52]. Further, effective use of AI techniques can enhance the energy efficiency and reliability of the CDC [67-68].

## 5.15 Containers

Existing systems are using Virtual Machines (VMs) for virtualization in fog and edge computing, which can be replaced with latest technology i.e. containers [69]. The utilization of containers can enable independence among execution of various applications which are running on same operating system. Further, the effective use of Docker based containers can reduce total cost of ownership.

## 5.16 Quantum Computing

The concept of quantum computing can be used to provide aid to various machine learning or deep learning techniques for the predication of resource requirement in fog and edge computing in advance [70]. Further, successful deployment of quantum based machine learning can help to manage fog and edge resources automatically and improve the resource utilization & energy efficiency.

## 5.17 Cloud Mining (Bitcoin)

The concept of cloud mining is introduced recently, which uses shared processing power to access remotely located cloud data center and produce the new bitcoins using computing power without managing the infrastructure [17]. Further, this concept can be extended towards fog and edge computing paradigms to optimize the utilization of computational or processing power.

## 5.18 Serverless Edge Computing

Serverless edge computing is new research area to explore where both Serverless and edge computing utilizing together to achieve a single objective [72]. Further, it provides the function as a service and executes individual functions on edge devices, which further saves energy consumption, reduces the latency & response time and improves the reliability while processing at closest edge device.

## 5.19 6G Technology

6G is the sixth generation and a successor of 5G technology to provision cellular data networks by increasing communication speed to a large extent [73]. The communication amongst IoT, fog and edge devices can be improved with the help of 6G technology [74]. During the execution of user requests, 6G can offer fast transmission while exchanging data among edge devices to reduce latency and response time.

## 5.20 Industry 4.0

The use of next generation technology in the area of manufacturing and industrial practices is increasing to automate the entire process for effective management of available resources within industry [44]. Smart industry can use AI or machine leaning based techniques to incorporate the automatic learning of the system, which can further improve the internal processes such as asset management or resource allocation [75].

## 5.21 Autoscaling

The utilization of Serverless computing or FaaS for fog and edge computing can satisfy the fluctuating demand of various IoT applications dynamically [1]. Further, the concept of autoscaling can improve the self-confirmation of resources, self-optimization of QoS parameters, self-protection from attacks and self-healing from occurrence of software, hardware or network faults without manual intervention [76].

## 6. Opportunities and Future Directions

Table 1 shows the various possible research opportunities and promising future directions for practitioners, researchers and academicians.

Table 1: Opportunities and Future Directions

| Relevant Issues and Paradigms | Opportunities and Future Directions |
|---|---|
| QoS and SLA | How to maintain the SLA and QoS at runtime during the execution of resources and workloads in fog and edge computing? |
| Energy Efficiency and Sustainability | How to deliver sustainable and energy efficient service using AI or machine leaning methods for fog/edge computing? |
| Resource Provisioning and Scheduling | How to provision fog and edge resources effectively for different IoT applications before actual scheduling of resources? |
| Fault Tolerance (Reliability) | How to maintain the reliability of service while delivering the sustainable service simultaneously? |
| Interlayer Communication | What is the impact of better interlayer communication among various layers of fog-edge computing model on QoS parameters? |
| Security and Privacy | How to maintain the privacy and security of fog/edge computing systems during the data processing of IoT applications? |
| Big Data Analytics | How to use AI or machine learning techniques for big data analytics for effective management and analysis of data? |
| Internet of Things | How to maintain the SLA during the processing of user requests in an efficient manner before their deadline? |
| Data Processing | What is the impact of data processing mechanisms on latency and response time? |
| Application Design | How to design new energy-efficient IoT applications for better utilization of fog/edge resources? |
| Serverless Computing | How to improve the scalability by using Serverless computing for fog/edge computing? |
| Blockchain | How Blockchain can be used to protect the data for IoT applications? |
| Software-Defined Network | How SDN can be used for fog/edge computing to reduce power consumption? |
| DL and AI | How to improve the deep learning and AI based resource management techniques? |
| Containers | How the utilization of containers for virtualizations can improve the QoS in fog/edge computing? |
| Quantum Computing | How to improve machine learning techniques using quantum computing? |
| Cloud Mining (Bitcoin) | How fog and edge computing can be used to optimize the utilization of computing or processing power? |
| Serverless Edge Computing | How FaaS can be used to improve the QoS in terms of power and utilization of resources? |
| 6G Technology | How 6G can be used to offer fast transmission while exchanging data among edge devices to reduce latency and response time? |
| Industry 4.0 | How to perform predicative analysis of industry assets using AI, fog and edge computing? |
| Autoscaling | How to provide effective autoscaling of resources to maintain the SLA and QoS at runtime for fog/edge computing? |

## 7. Summary and Conclusions

The use of IoT applications are increasing day by day and producing lot of data in seconds. Cloud platform is effective in data management dynamically but latest IoT applications needs to process data with minimum latency and response time. In this chapter, a manifesto for modern fog and edge computing systems is presented to evaluate the ongoing research in this field.

Further, type of architectures and applications for fog and edge computing are presented. Finally, research opportunities and promising future directions are highlighted.

## List of Abbreviations

**CDC:** Cloud Data Centres

**IoT:** Internet of Things

**6G:** 6th Generation

**VM:** Virtual Machines

**SLA:** Service Level Agreements

**QoS:** Quality of Service

**AI:** Artificial Intelligence

**DoS:** Denial of Service

**DDoS:** Distributed-DoS

**R2L:** Remote to Local

**U2R:** User to Root

**FaaS:** Function as a Service

## About the Author

**Sukhpal Singh Gill** is a Lecturer (Assistant Professor) in Cloud Computing at School of Electronic Engineering and Computer Science, Queen Mary University of London, UK. Prior to this, Dr. Gill has held positions as a Research Associate at the School of Computing and Communications, Lancaster University, UK and also as a Postdoctoral Research Fellow at CLOUDS Laboratory, The University of Melbourne, Australia. Dr. Gill is serving as an Associate Editor in Wiley ETT and IET Networks Journal. His research interests include Cloud Computing, Fog Computing, Software Engineering, Internet of Things and Healthcare. For further information, please visit http://www.ssgill.me